# Quantum Inverse Fast Fourier Transform

S Mayank Roy
*Dept. of Electrical and Electronics Engineering*
*Rajalakshmi Engineering College*
Chennai, India
mayankroy0987@gmail.com

V Devi Maheswaran
*Dept. of Electrical and Electronics Engineering*
*Rajalakshmi Engineering College*
Chennai, India
devimaheswaran.v@gmail.com

*Abstract*— **In this paper, an algorithm for Quantum Inverse Fast Fourier Transform (QIFFT) is developed to work for quantum data. Analogous to a classical discrete signal, a quantum signal can be represented in Dirac notation, application of QIFFT is a tensor transformation from frequency domain to time domain. If the tensors are merely complex entries, then we get the classical scenario. We have included the complete formulation of QIFFT algorithm from the classical model and have included butterfly diagram. QIFFT outperforms regular inversion of Quantum Fourier Transform (QFT) in terms of computational complexity, quantum parallelism and improved versatility.**

## I. INTRODUCTION

A Fourier transform is an integral transformation method, which takes input as a function and gives output of its decomposed frequencies and amplitudes. This idea was first given by Joseph Fourier in 1822 that a complex function can be broken down into a series of sines such that it is a Lebesgue integrable function. This was further modified into the version we know today [1], [2]. In digital signal processing, an important result obtained from Fourier transform is called Discrete Fourier Transform (DFT), where the transformation is done at discrete intervals for a discrete input signal [3], [4]. For a discrete time signal:

$$\{x_v\}, v = 0, 1, \dots, N-1 \tag{1}$$

The DFT is given by:

$$X(k) = \sum_{n=0}^{N-1} x(n) \exp\left(-\frac{i2\pi nk}{N}\right) \tag{2}$$

Where 'i' denotes the complex number $\sqrt{-1}$. From the Fourier inversion theorem [5], it is noted that an inverse for every DFT operation can be obtained. The Fourier inversion theorem is mathematically represented as:

$$f(x) = \int_{-\infty}^{\infty} \hat{f}(\omega) \exp(i2\pi\omega x)\, d\omega \tag{3}$$

The expression for Inverse Discrete Fourier Transform [6] when passed through inversion theorem is given as:

$$x(n) = \frac{1}{N}\sum_{k=0}^{N-1} X(K) \exp\left(\frac{i2\pi nk}{N}\right) \tag{4}$$

A common procedure used for solving DFT is called the Fast Fourier Transform (FFT) [7], [8]. In a work by Asaka et.al. [9], a novel algorithm for Quantum Fast Fourier Transform (QFFT) was proposed with a computational complexity of $O(Nlog_2 N)$ which outperforms Quantum Fourier Transform (QFT) with a computational complexity of $O(N^2)$. At super low temperatures, by making use of quantum entanglement, we also enable quantum parallelism, which allows use to perform simultaneous operations. Another advantage of using quantum mechanics in the context of signals is data storage efficiency and data security [10]. As discussed in [9], a procedure for QFFT is obtained by employing FFT in the quantum scenario through tensor transformation-based basis encoding. In this paper, a procedure has been laid out for the inverse process, which we would call as Quantum Inverse Fast Fourier Transform (QIFFT) through a similar encoding process.

## II. MATHEMATICAL FORMULATION OF QIFFT

Throughout this paper, Dirac notation will be adopted. For a vector space $\mathbb{V}$, the vectors will be denoted by, $|v\rangle$. An in-depth treatment of Dirac notation and quantum mechanics can be found in [11]. Let $\{x_i\}$ represent the input quantum data sequence and $\{X_j\}$ represent the output quantum data sequence. The forward transformation through tensor product can be represented as:

$$\otimes |x_i\rangle \rightarrow \otimes |X_j\rangle \tag{5}$$

Then, we re-write IDFT equations accordingly as:

$$|x_i\rangle = \frac{1}{N}\sum_{K=0}^{N-1} |X_k\rangle W_N^{nk} \tag{6}$$

Where, $W_N^k = \exp(-2\pi ik/N)$. The above equation is for IDFT through amplitude encoding, for basis encoding, we need to re-write it as:

$$\otimes |X_j\rangle \rightarrow \frac{1}{N} \otimes |x_i\rangle \tag{7}$$

In this paper, we work with only Decimation-In-Time (DIT) Fast Fourier Transform (FFT) procedure and develop the algorithm for QIFFT. Assume a data sequence $\{X_j\}$ in the frequency domain. By taking its tensor product, we get:

$$\otimes |X_j\rangle \tag{8}$$

In DIT-FFT, we decompose this product into odd and even parts, doing so:

$$|x_k\rangle' = |F_k^{(*,n-1,0)} + W_N^k F_k^{(*,n-1,1)}\rangle \tag{9}$$

$$\left|x_{k+\frac{N}{2}}\right\rangle' = \left| F_k^{(*,n-1,0)} + W_N^k F_k^{(*,n-1,1)} \right\rangle \tag{10}$$

Where $0 \leq k \leq N/2 - 1$, The * denotes the complex conjugate taken at the input. Here, b denotes the batch from which the previous coefficient comes, $F_k^{(*,n-1,b)}$ represents Fourier co-efficient. It is evaluated by:

$$F_k^{(*,n-1,b)} = \sum_{j}^{N-1} W_N^{2ik} |X_{2j+b}\rangle \tag{11}$$

To obtain the final tensors, we need to modify the equations as:

$$|x_k\rangle = \frac{1}{N}(|x_k\rangle')^* \quad (12)$$

$$|x_{k+N/2}\rangle = \frac{1}{N}(|x_{k+N/2}\rangle')^* \quad (13)$$

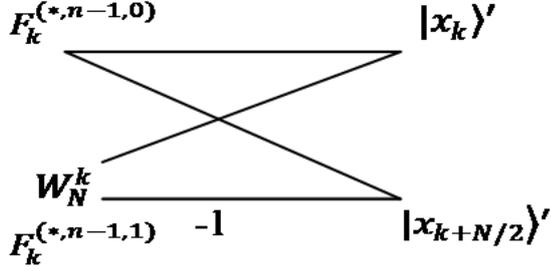

Fig. 1. General Butterfly Diagram for QIFFT

Equations (9)-(13) is used to develop the abstract mathematical background of the final step of QIFFT. To solve FFT, a simpler approach called butterfly diagram is often employed, whose formulations are discussed in [4]. Algorithmic procedure for solving QIFFT is discussed in the subsequent section. Fig. 1 above shows the butterfly diagram for QIFFT developed for equations (9)-(13). Our final step is to obtain the recursive equations for the Fourier coefficients. For iteration k, at n-m step, we get:

$$\left|F_k^{(*,n-m,b)}\right\rangle = \left|F_k^{(*,n-m-1,b)} + W_{N/2^m}^k F_k^{(*,n-m-1,b+2^m)}\right\rangle \quad (14)$$

$$\left|F_{k+\frac{N}{2^{m+1}}}^{(*,n-m,b)}\right\rangle = \left|F_k^{(*,n-m-1,b)} + W_{N/2^m}^k F_k^{(*,n-m-1,b+2^m)}\right\rangle \quad (15)$$

The input Fourier coefficients are the complex conjugates of the data sequence.

### III. ALGORITHM FOR QIFFT

Two vector operations which are used frequently in QIFFT are vector addition and scalar multiplication. Given kets $|a\rangle$ and $|b\rangle$, their vector sum is:

$$|a\rangle + |b\rangle = |a+b\rangle \quad (16)$$

Now let's assume we have a ket $|a\rangle$ and a scalar $\Psi$, then the scalar product is given by:

$$\Psi|a\rangle = |\Psi a\rangle \quad (17)$$

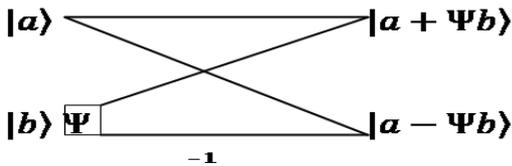

Fig. 2. Butterfly Diagram for the General Operation With Kets

Equations (16) and (17) can be used to construct a generalized operating butterfly diagram which is shown above in Fig. 2. Using the techniques discussed so far, we can now develop the algorithm to take QIFFT.

1. Take complex conjugate of the input quantum data sequence.

2. Perform DIT-QFFT procedure successively.

3. Take the complex conjugate and divide the result by length of the data sequence.

At the end of this process, we would have obtained output quantum data sequence in time domain.

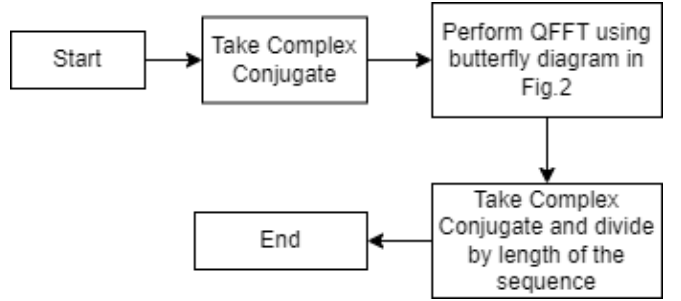

Fig. 3. Flowchart of QIFFT algorithm

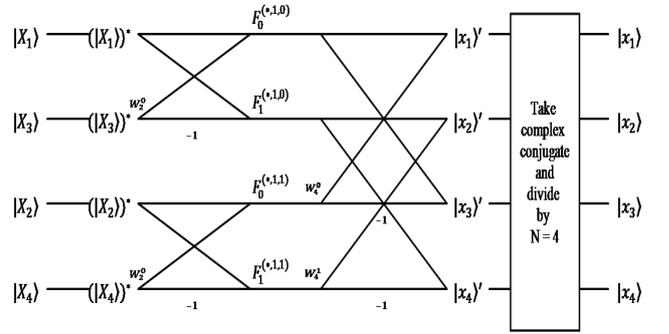

Fig. 4. QIFFT structure for a data sequence of length 4

Fig. 3 shows the flowchart of QIFFT algorithm. Fig. 4 here shows the procedure for QIFFT for a data sequence of length 4. It is to be noted that when QIFFT is performed in DIT procedure, the input sequence must be given in bit reversal order.

### IV. CONCLUSION

In this work, we have developed a procedure to take the inverse for the Quantum Fast Fourier Transform and developed an algorithmic procedure to solve it using butterfly diagram. QIFFT may find uses in the future as inverses of data from frequency domain to time domain are often taken at the receive end to either decode the data or check the data integrity. This papers' main aim is to only show the possible methods to take generalized FFT inverses using quantum mechanics, and we have only used one procedure to obtain the results to act as a proof of concept. As IBM, Microsoft and others develop quantum computers with greater number of qubits (and hence, the greater quantum processing power), it can be expected in the foreseeable future to witness a boom in quantum algorithms applied to various fields like signal processing, machine learning, cryptography and more.